\documentclass[preprint,12pt]{elsarticle}

\usepackage{amsmath}
\usepackage{amssymb}
\usepackage{dsfont}
\usepackage{color}
\usepackage{graphicx} 

\usepackage{standalone}
\usepackage{subfig}
\usepackage{multirow}
\usepackage{morefloats}
\usepackage{pgfplots}
\usetikzlibrary{plotmarks}
\pgfplotsset{compat=newest}
\usepackage{tikz}
\usepackage{sty/relinput}
\usepackage{ifdraft}

\usepackage{filecontents}
\usepackage[Publish,Externaldir=pictures/external_pics/]{sty/externaltikz}

\journal{Applied Mathematical Modelling}

\begin{document}

\begin{frontmatter}
\title{Experimental Validation of a Microscopic Ellipsoidal Particle Model Immersed in Fluid Flow}

 \author[a1]{A. Meurer}
 \author[a2]{A. Weber}
 \author[a2] {H.-J. Bart} 
  \author[a1]{A. Klar}
 
 \address[a1]{Technische Universit\"at Kaiserslautern, Department of Mathematics, Erwin-Schr\"odinger-Stra{\ss}e, 67663 Kaiserslautern, Germany 
  \{meurer,klar\}@mathematik.uni-kl.de}
 \address[a2]{Technische Universit\"at Kaiserslautern, Department of Process Engineering, Erwin-Schr\"odinger-Stra{\ss}e, 67663 Kaiserslautern, Germany  \{andreas.weber,bart\}@mv.uni-kl.de}

\begin{abstract}
A simulation of ellipsoidal particles inside a fluid flow is presented and validated from our own lab size experiments where data has been recorded using a camera set-up and post-processing of the pictures. The model is based on Jeffery's equation, where the orientation and position of the particles are influenced by the surrounding fluid. Additionally, particle-particle interaction and particle-wall interaction are taken into account.
\end{abstract}


\end{frontmatter}

\section{Introduction}
In many industrial applications the flow behavior of particles can be an important process parameter. Especially when it comes to non-spherical particles, the prediction of movement properties is difficult. Particle-particle and particle-wall collisions must be regarded as fundamentally different than it is the case with spherical particles. While spherical particles have a isotropic behavior when it comes to drag force and dimension, it allows for simple calculation of forces acting. This is not valid any more when observing deformed particles - forces always have to be calculated with respect to the actual dimensions/directions of the considered particle \cite{Senchenko.2006,Zhou.2011}.\\
Mathematically, the movement of ellipsoidal particles, can by described by a system of stochastic differential equations based on Newtonian laws of mechanics. To model the movement suspended in an incompressible flow, one can use the model of Jeffery \cite{jeffery1922motion,junk2007new,walter2010stochastic}, which describes the forces acting on the particles, depending on the surrounding fluid. Additionally, the particle-particle interaction of the ellipses are described via pairwise interaction potentials and a random force. The potentials we use are common in the literature of polymers \cite{berardi1995generalized,berne1972gaussian,cleaver1996extension,everaers2003interaction,gay1981modification,perram1996ellipsoid}, where the shape of the ellipses are modeled with the help of Gaussian type functions. This leads to a model similar to the one described in \cite{coffey1996langevin,han2006brownian,tavakol2015dispersion,walter2010stochastic}. For different level of descriptions for this particle model, see \cite{borsche2015mean}.\\
Based on this ellipsoidal particles the model could be expanded to a wider range of industrial applications such as simulation of bubbles inside chemical reactors or treatment/classification of grains . Often these problems are calculated based on simple spherical particle models and could thereby profit from this new model approach \cite{Ottino.2001,Jakobsen.2005}.
The paper is organized as: In section \ref{sec:experiment}, we describe the lab experiment, where a channel was filled with water and ellipsoidal particles. A camera system has been used to evaluate different parameters of the particle flow. With a suitable image analysis software we were able to determine movement speed, orientation and residence density of the considered particles.\\
In section \ref{sec:cfd}, a CFD simulation of the fluid flow inside the channel is investigated and needed for the ellipsoidal particle model. The mathematical model is described in section \ref{sec:model} and the needed parameters for the comparison to the lab experiment is considered. This enables us to compare the model to the experimental data in section \ref{sec:comparison}.


\section{Experiment}
\label{sec:experiment}
\begin{figure}[h]
\centering
\includegraphics[width=0.49\textwidth]{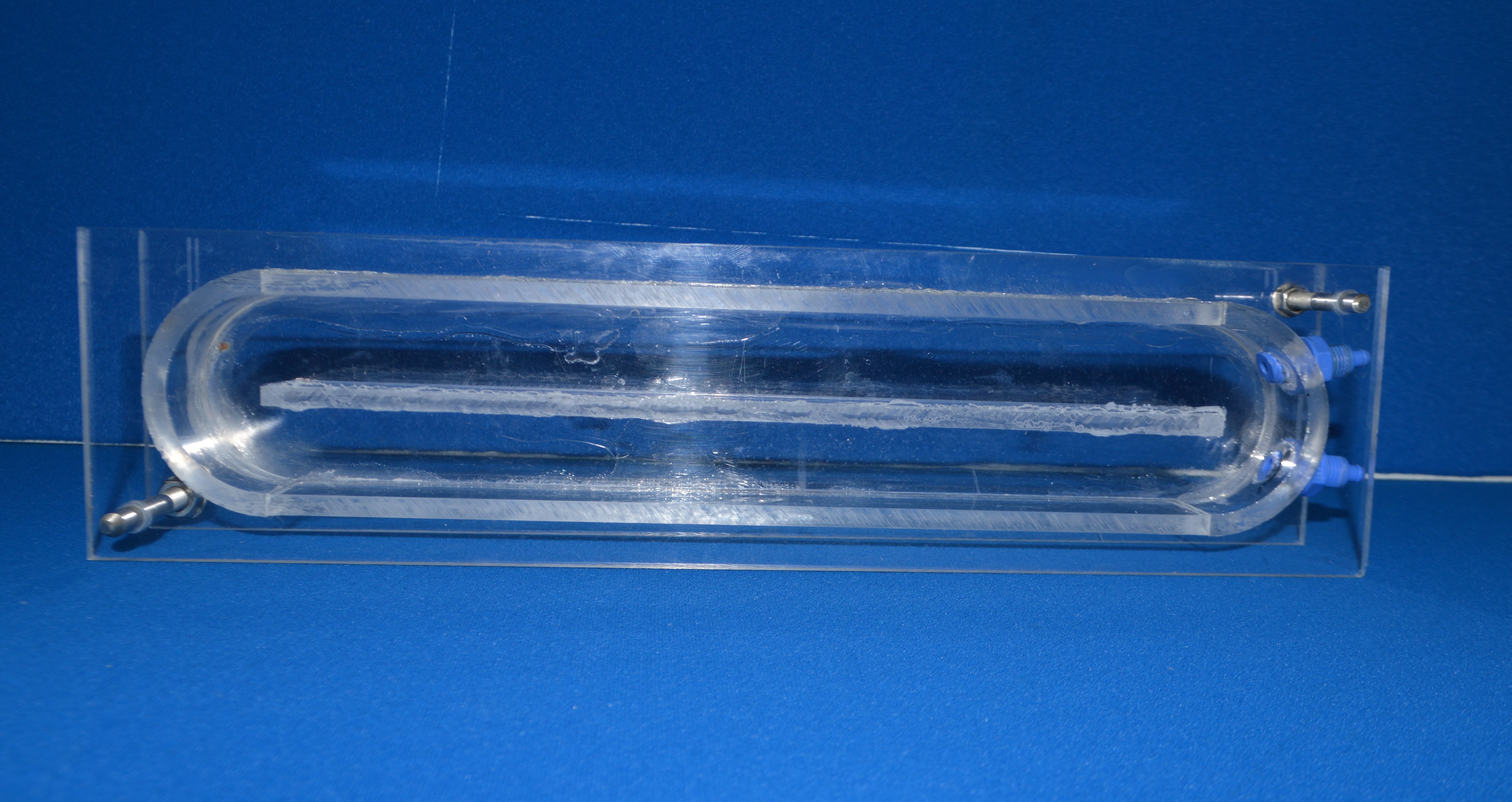}
\includegraphics[trim = 0mm 30mm 0mm 0mm, clip, width=0.49\textwidth]{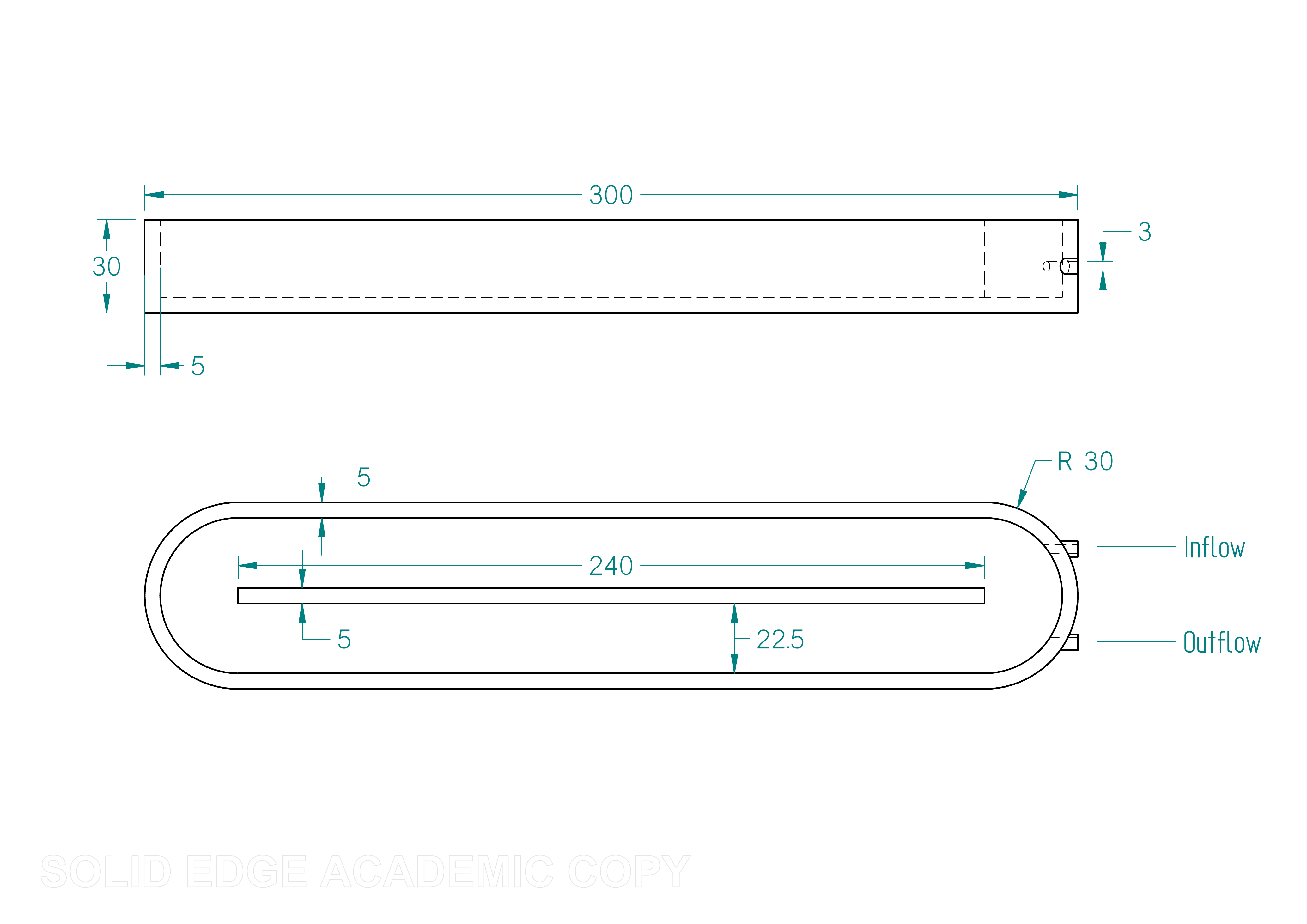}
\caption{Picture of the acrylic glass channel on the left and the corresponding measures (in mm) on the right.}
\label{fig:kanal}
\end{figure}

We consider an acrylic glass channel with measures as shown in Figure \ref{fig:kanal}. The channel is completely filled with reverse osmosis water, air bubbles at the lid were carefully removed.
In- and outflow are connected to a pumping system which gives a constant flow of $\dot{V}=66.24~\frac{l}{h}$. At the inflow cross section of $A = 9.2~mm^2$ an average fluid velocity of $u = 2~\frac{m}{s}$ is reached. The water flows in at the top channel and forms a circular stream in counter-clockwise direction.
\begin{figure}[h]
	\centering
	\includegraphics[width=0.49\textwidth]{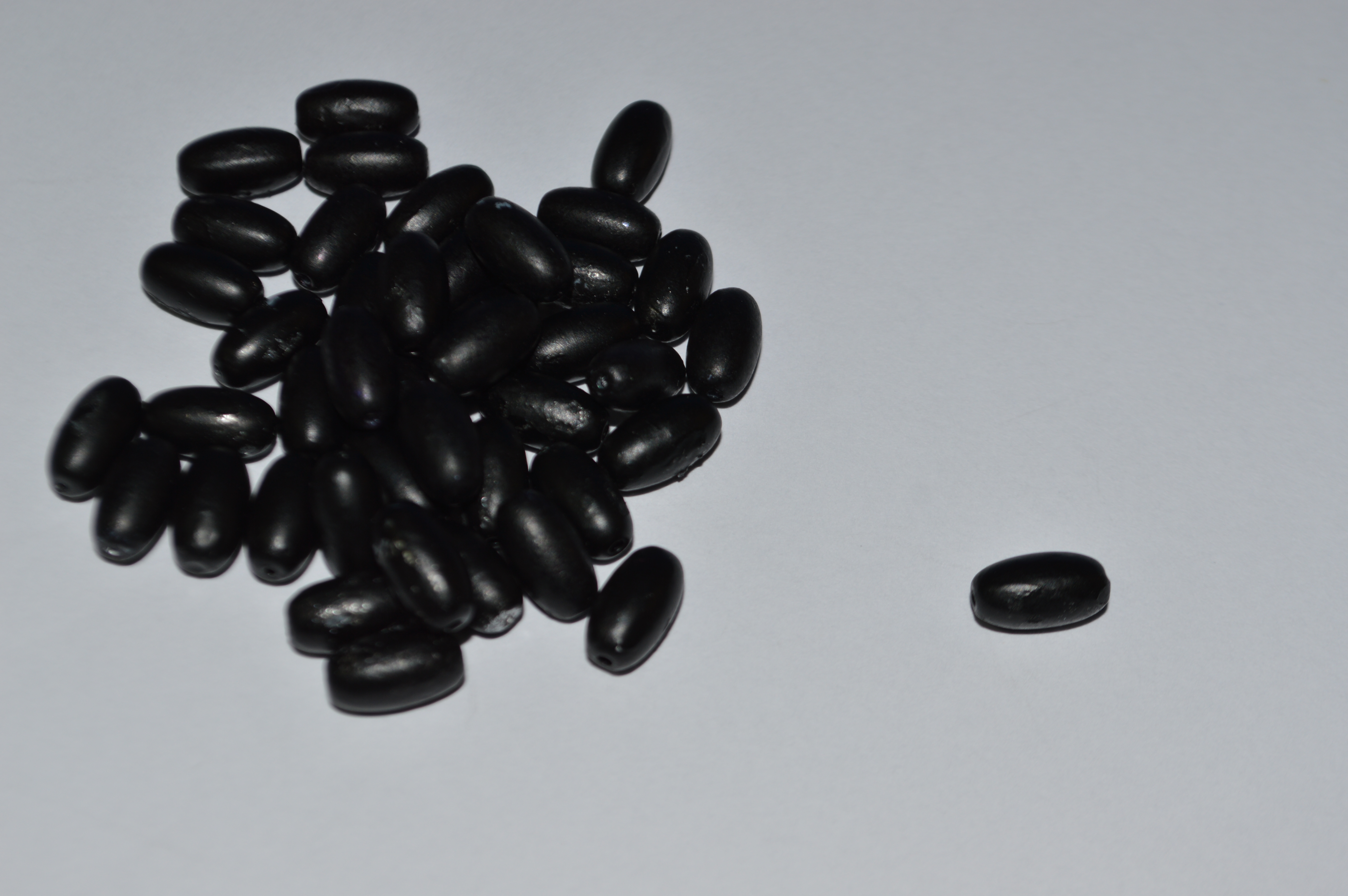}
	\includegraphics[width=0.49\textwidth]{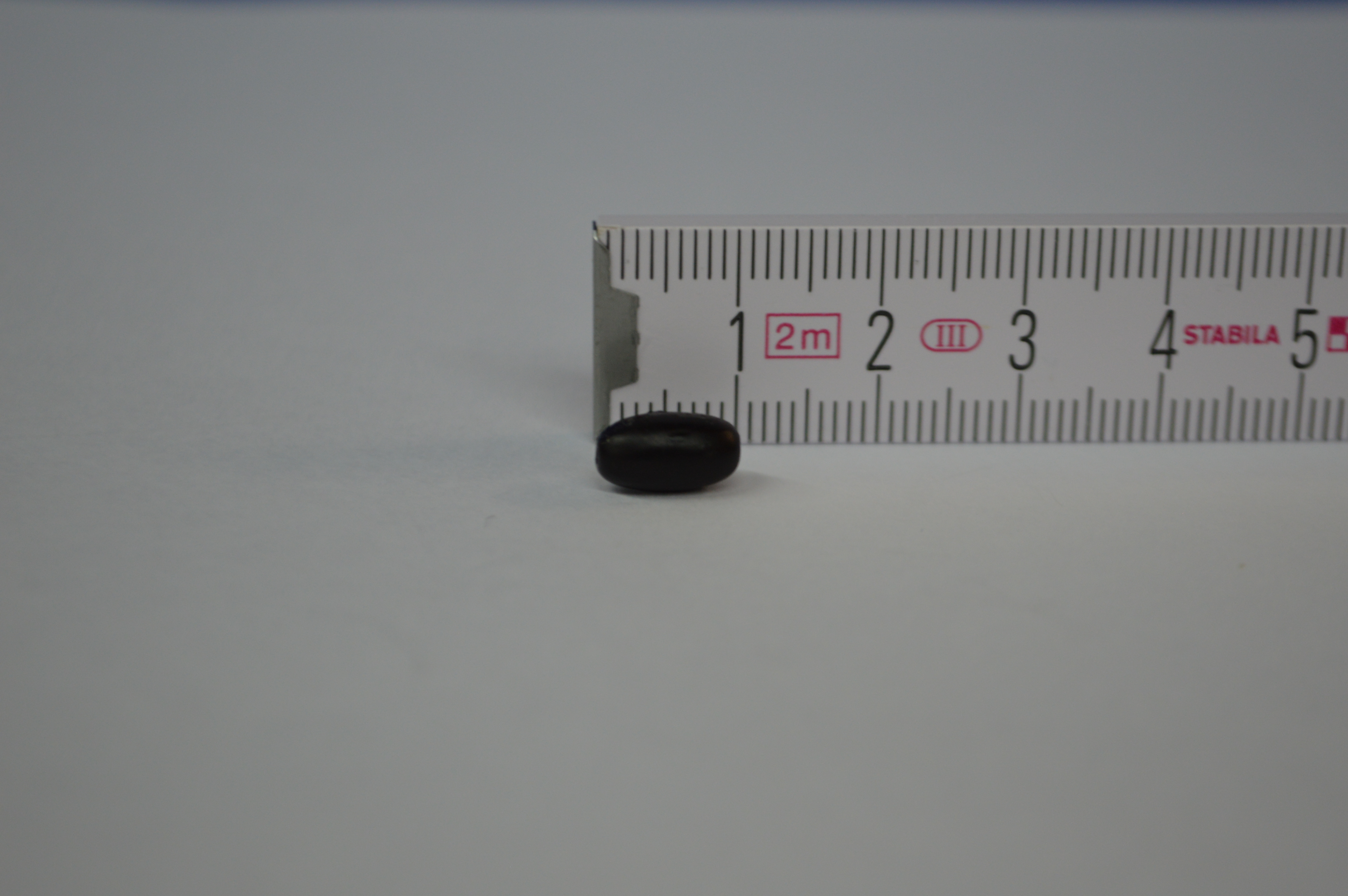}
	\caption{The particles which are used for the experiment.}
	\label{fig:partikel}
\end{figure}
The particles added to the flow resemble prolate rotational ellipsoids with a major radius of $L=4.9~mm$ and minor radius of $D=2.5~mm$ as shown in Figure \ref{fig:partikel}. Due to their density of $\rho=0.95~\frac{g}{cm^3}$ the particles swim slightly inside the water with $\rho=1~\frac{g}{cm^3}$. Particles have been tested individually and in an ensemble of $N=25$ particles to detect the orientation and position inside the channel. Since the flow inside the pump and channel can not be switched on instantaneously, we start the detection of  particles when the steady-state flow is reached. Note that in this case the particles are already moving and can not be held at a particular position. We detect the particle motion for $T=10~s$ with a frame rate of $fps = 25~s^{-1}$.
\subsection{Post-processing}
In the post-processing step we use the resulting pictures to analyze the particle information with the help of the program ImageJ \cite{imagej}. Since the particles have a rather dark color, it is possible to separate them with a threshold binarization technique. In a second step a watershed algorithm separates particles which are in direct contact and apart from that would be recognized as one larger particle. At last ImageJ measures the position, orientation, size and boundary line of each particle. This embodies the major and minor axis lengths and the angle of the major axis compared to the picture coordinate system. In total we obtain the spatial and directional distribution and the velocity of the particles.
\begin{figure}[h]
	\centering
	\includegraphics[width=0.25\textwidth]{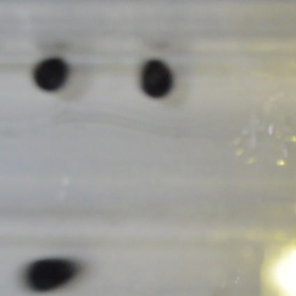}
	\includegraphics[width=0.25\textwidth]{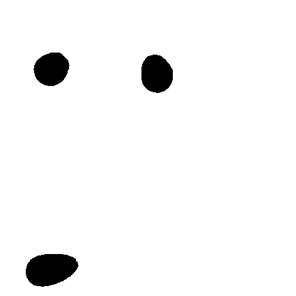}
	\includegraphics[width=0.25\textwidth]{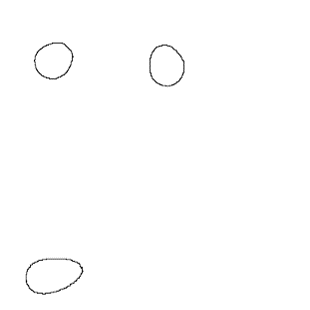}
	\caption{Stepwise post-processing of a picture in ImageJ}
	\label{fig:imageJ}
\end{figure}
Another approach persists in utilizing the standard deviation of the pictures. An area of the channel, where it is not possible to find a particle would always show the same color/brightness on every picture. Particles will rise the deviation of the brightness the more often they pass a pixel. Hence the deviation of brightness values also gives the spatial distribution of the particles. This can also be done with or without binarizing the pictures at first.
\begin{figure}[h]
	\centering	
	\includegraphics[width=0.45\textwidth]{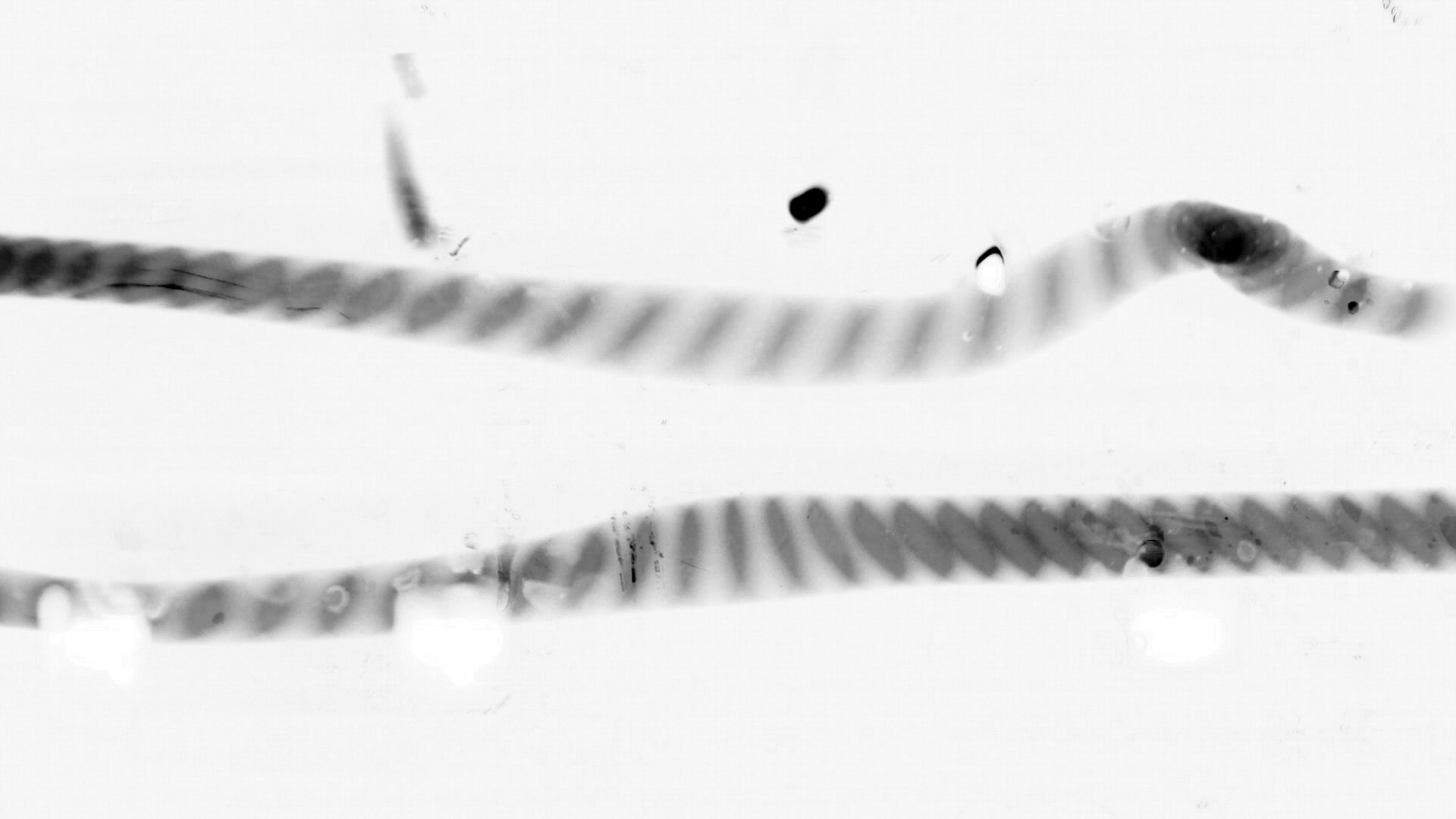} 
	\includegraphics[width=0.45\textwidth]{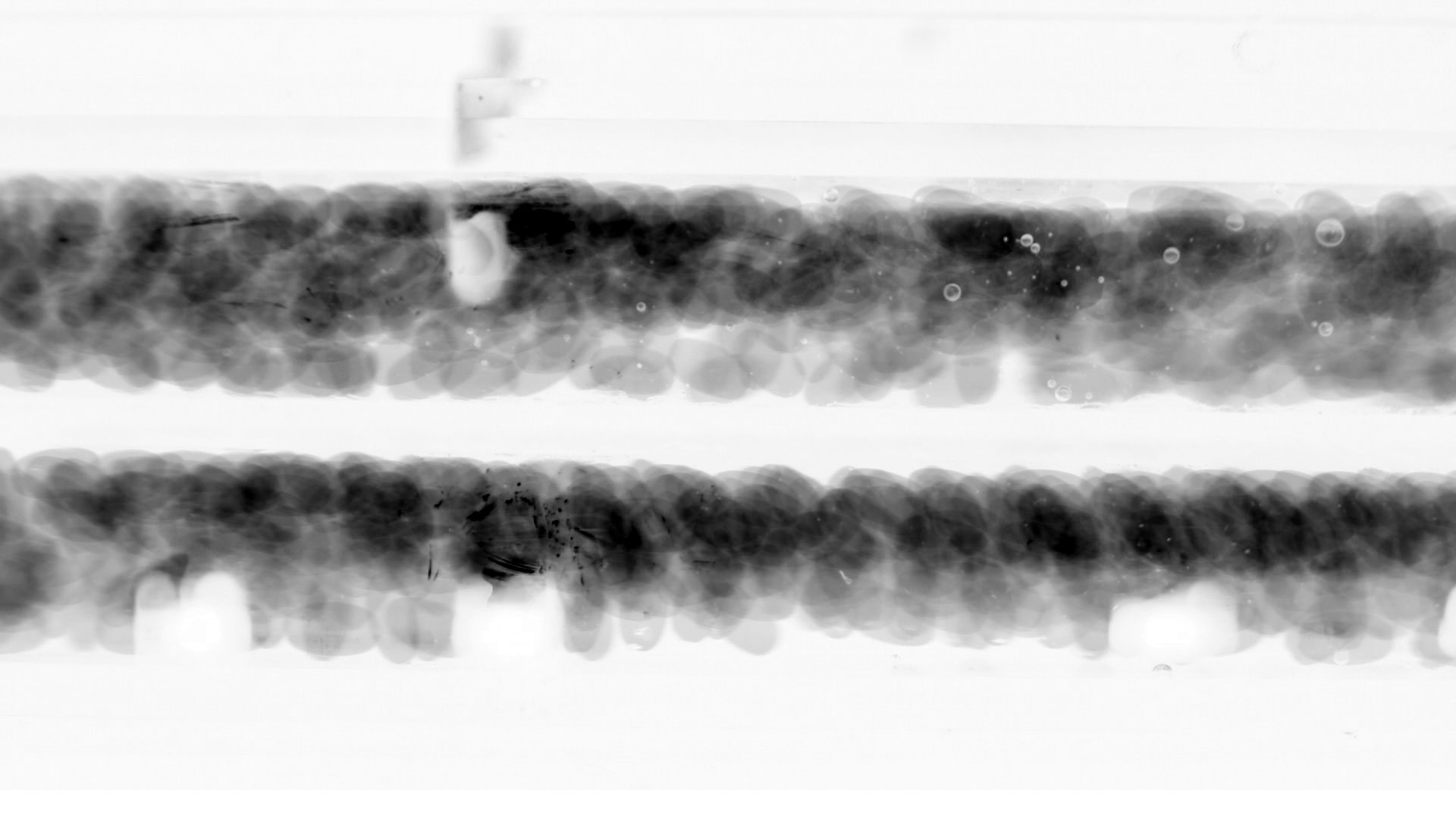}
	\caption{Brightness deviation of pictures from single and multiple particle experiments}
	\label{fig:deviation_images}
\end{figure}\\
Figure \ref{fig:deviation_images} shows resulting deviation pictures, where darker areas stand for higher residence probability of a particle. On the left picture, note the clearly visible single particle trajectory and position of wall contact. Also the particle velocity can be deduced from the pattern of the trajectory. Here the particle is much faster in the upper channel than it is in the lower one.

The right picture was calculated from the multiple particle experiment. It is clearly visible that particles are not scattered uniformly but have a higher probability to be found near the walls of the channel. In the upper channel they are more concentrated at the outer wall while in the lower channel they appear to be more present at the inner wall. The lighter spots in the picture are a result of measurement errors, that originate from light reflections on the lid of the channel.
\section{CFD simulation}
\label{sec:cfd}
\begin{figure}[h]
	\centering
	\includegraphics[width=0.8\textwidth]{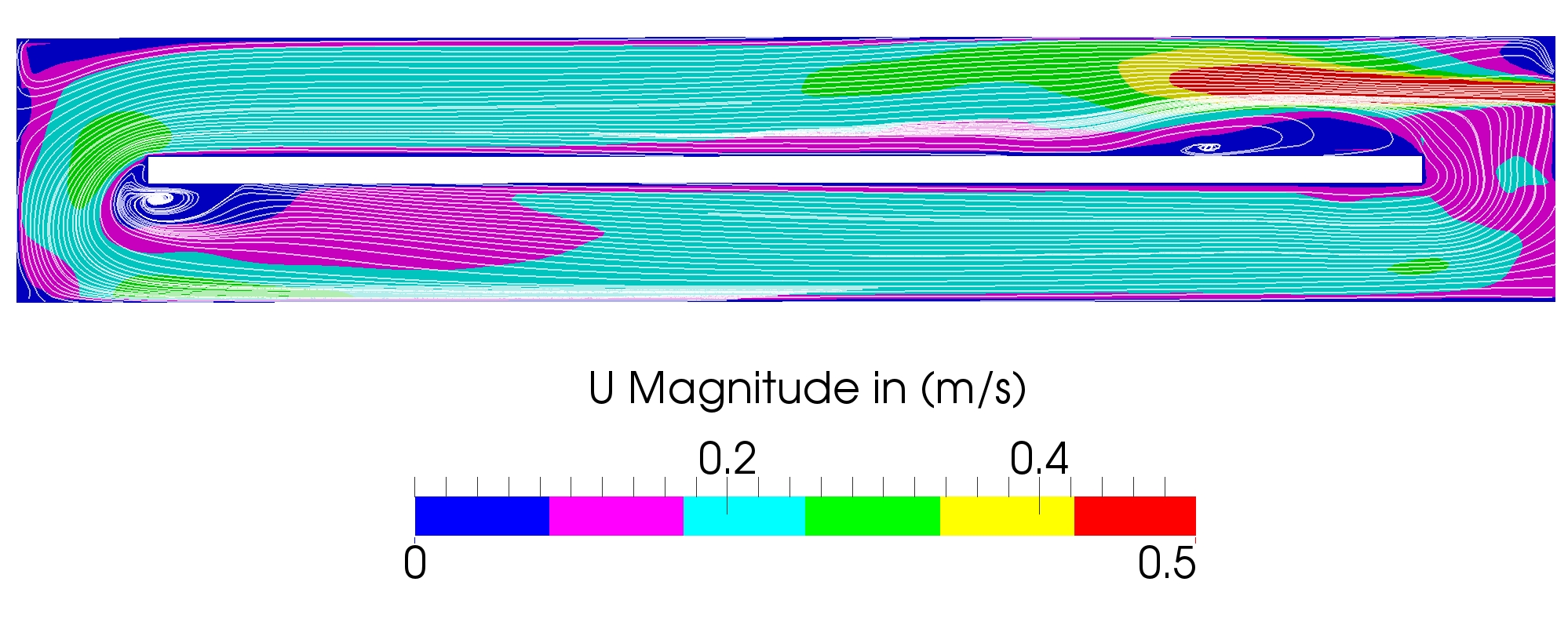}
	\caption{CFD simulation results showing fluid velocity and streamlines.}
	\label{fig:CFD}
\end{figure}

A CFD simulation for incompressible laminar flow was chosen to calculate the flow pattern inside the channel. This fluid velocity field was then used for the particle simulation and evaluation of experimental data. For simplicities sake, the CFD domain was chosen rectangular, while the experimental has rounded corners. This simplification was made due to the particle model having trouble with curved edges. Nevertheless, measurements have been solely made on the middle part of the channel, the corners have not been part of any experimental evaluation.
The measures are considered in $cm$. Hence our fluid domain is given by
\begin{align*}
\Omega_F=[0~29]\times[0~5]\times[0~2.5]\setminus[2.5~26.5]\times[2.25~2.75]\times[0~2.5].
\end{align*}
 For the boundary conditions we choose
 \begin{itemize}
\item Inflow velocity of $2m/s$ in $\Omega_{IN}=\{29\}\times[3.7~0.4]\times[1.1~1.4]$.
\item Free outflow at $\Omega_{OUT}=\{29\}\times[0~2.25]\times[0~2.25]$.
\item No-slip condition for the other boundaries.
\end{itemize}
Transport properties were set to standard values of pure water, namely incompressible fluid with density $\rho=1000~\frac{kg}{m^3}$ and viscosity $\nu=10^{-6}\frac{m^2}{s}$.\\
The typical flow pattern is shown in Figure \ref{fig:CFD}. The high inflow velocity causes the flow to fluctuate inside the upper channel. The flow data used in the particle simulation was therefore smoothed by averaging the velocity field over $\Delta t = 20~s$.
Since the inflow velocity is rather high, a clearly developed eddy shows up in the upper channel. In this area the fluid can move against the overall counter-clockwise direction eventually dragging particles with it. This behavior could be confirmed in the experiment (and simulation). However, the lower channel shows a more smooth velocity profile, where only a small eddy is present directly behind the bend. This moderate velocities also simplified the measurements in the lower channel. The outflow in the lower right corner of the channel shows no significant influence on the flow pattern.

Since the CFD simulation was solved for the 3D case, data had to be converted to the 2D model. This was done by mapping a plane through the middle of the computational domain. Fluid velocities in z-direction were small enough to be neglected, thus the resulting 2D velocity field could still reflect the flow pattern.

\section{The Model}
\label{sec:model}
To describe the movement of the ellipsoidal particles suspended in an incompressible fluid, we consider a microscopic Langevin-type model as in \cite{walter2010stochastic}. Since we just detect the particle orientation in the experiment with one camera, we assume that the height of the particles inside the channel is not relevant for our test case. Therefore we assume that the particles are two dimensional. The interaction of the fluid with the ellipses is described by a Jeffery's type term \cite{jeffery1922motion,junk2007new} and the particle interaction is given by a many-particle interaction potential similar to \cite{berne1972gaussian}. Each particle is described by its position $r_t\in\mathbb{R}^2$, velocity $v_t\in\mathbb{R}^2$, orientation angle $\theta_t\in[0,~2\pi)$ and angular velocity $\omega_t\in\mathbb{R}$. The angle $\theta_t$ describes the relative angle between the horizontal axis and the main axis of the particle, such that $\theta_t=0$ corresponds to the orientation $(1,~0)^T$. Then the equation of motion for $N$ particles $i=1,\ldots,N$ are
\begin{align}\label{eq:micro} 
\left\{\begin{aligned}
dr_t^i &= v_t^i dt\\
dv_t^i &= \gamma(u-v_t^i)dt-\frac{1}{m}\frac{1}{N}\sum_{i\neq j} \nabla_{r_i}U(r_t^i,r_t^j,\theta_t^i,\theta_t^j)dt\\
&\hspace*{10em} - (A^2/2)v_t^idt+A dW_t^{A,i}\\
d\theta_t^i &= \omega_t^idt\\
d\omega_t^i &= \bar{\gamma}(g(\theta_t^i,u)-\omega_t^i)dt-\frac{1}{I_c}\frac{1}{N}\sum_{i\neq j} \nabla_{\theta_i}U(r_t^i,r_t^j,\theta_t^i,\theta_t^j)dt\\
&\hspace*{10em} - (B^2/2)\omega_t^idt+Bd W_t^{B,i},
\end{aligned}\right.
\end{align}
with appropriate initial conditions.
Here $u$ is the velocity of a surrounding fluid. For the surrounding fluid, we assume that the influence of the particle to the fluid is neglectable such that we use a stationary fluid in the simulation. The function $g(\theta,u)$ is given by
\begin{align*}
g(\theta,u)&=\frac{1}{2}\text{curl}(u)+\lambda\left(
\begin{array}{r}
-\sin\theta\\
\cos\theta
\end{array} \right)^\top\left(\frac{1}{2}(\nabla u+\nabla u^\top)\right)\left(
\begin{array}{c}
\cos\theta\\
\sin\theta
\end{array} \right).
\end{align*}
The first terms on the right hand side of  the velocity and angular velocity equations describe
the relaxation  of the particles to  the velocity of the fluid and to the rotation resulting from the velocity field, respectively.
The speed of relaxation is determined by the friction parameters $\gamma$ and $\bar{\gamma}$. 
The second term models the repulsive interaction between the particles.
To model the interaction between two ellipsoidal particles, there exist many different potentials \cite{berardi1995generalized,berne1972gaussian,cleaver1996extension,everaers2003interaction,gay1981modification,perram1996ellipsoid}.
We use the soft potential as proposed by Berne \cite{berne1972gaussian}.
It is obtained by overlapping two ellipsoidal Gaussians representing the mutual repulsion of two particles. 
This leads to
\begin{align*}
\tilde{U}(r,\bar r,\theta,\bar \theta)&=a(\theta,\bar \theta)
\exp\left(-\left(\bar r-r\right)\left(\gamma (\theta)+\gamma(\bar \theta) \right)^{-1}\left(\bar r-r\right)\right),
\end{align*}
where $a$ and $\gamma$ are defined by
\begin{gather*}
a(\theta,\bar \theta)=\epsilon_0\left(1-\lambda^2
(
\eta(\theta)\cdot\eta(\bar\theta))^2\right)^{-\frac{1}{2}},
\qquad \eta(\theta)=(\cos\theta, \sin\theta)^\top,\\
\gamma (\theta) =\left(l^2-d^2\right)
\eta(\theta)\otimes
\eta(\bar\theta)
+d^2\mathds{1},\qquad
\lambda=\frac{l^2-d^2}{l^2+d^2}.
\end{gather*}
Here, $l=2L$ and $d=2D$ where $L$ is the major and the $D$ the minor radius of the particle. 
The parameter $\epsilon_0$ models the strength of the potential.
To have compact support we slightly modify the potential and define
\begin{align}
U(r,\bar r, \theta,\bar \theta)&=a(\theta, \bar \theta)
\exp\left(-\frac{\left(\bar r-r\right)\left(\gamma(\theta)+\gamma(\bar \theta)\right)^{-1}\left(\bar r-r\right)}{1-\left(\bar r-r\right)\left(\gamma(\theta)+\gamma(\bar\theta)\right)^{-1}\left(\bar r-r\right)}\right).
\label{eq:interaction}
\end{align}
The parameters $m$ and $I_c$ are the mass and the moment of inertia of the particles.
Furthermore, $A,B$ are non-negative diffusion constants and $W^{A,i},W^{B,i}$ are independent standard Brownian motions. 
\subsection{Numerical Set Up}
\begin{figure}[htbp]
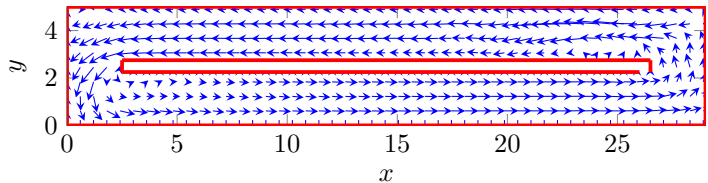

   \centering
   \externaltikz{ExternalTikzFigure_vecfeld_exp}{
   	\relinput{pictures/tikz/vecfeld_exp/vecfeld_exp.tex}
   }
   \caption{Experiment: velocity field of the surrounding fluid is given by the blue arrows, and the wall boundaries are indicated by the red line.}
   \label{fig:vecfeld_exp}
\end{figure}

We consider the length of the channel in $cm$, such that we have the following domain
\begin{align*}
\Omega=[0~29]\times[0~5]\setminus[2.5~26.5]\times[2.25~2.75].
\end{align*}
To include wall boundaries, we insert ghost particles with distance $l/2$ onto the boundaries, where their orientations lie parallel to the respective wall. The interaction potential is the same as before \eqref{eq:interaction}. For the interaction of the boundary particles with the inner particles, we increase the value of $\epsilon_0$ by a factor of 10. For the surrounding fluid we us the results of the CFD simulation as described in Section \ref{sec:cfd}.
To use the resulting velocity field in the two-dimensional set up, we choose the time-averaged velocity field of the middle plane in $z$-direction (compare Figure \ref{fig:vecfeld_exp}).
\\
 Then we set the following parameters according to the experiment 
\begin{align*}
l&=0.49~cm, & d&=0.25~cm, & T=10~s.
\end{align*} 
The mass $m$ and moment of inertia $I_c$ for the ellipsoidal particles are given by
\begin{align*}
m&=\rho V=\rho\frac{4}{3}\pi LD^2=0.97~g,\\
I_c&=\frac{m}{5}(L^2+D^2)=0.25~gcm^2.
\end{align*}
The friction parameters for translational motion $\gamma_t$ and rotational motion $\gamma_r$ are only given into the direction of the main axis of the particle \cite{sun2008langevin,norisuye1979wormlike}. They have the following form
\begin{align*}
T_t^L&=16\pi\bar{\eta}\frac{L^2-D^2}{(2L^2-D^2)S-2L},\\
T_t^D&=32\pi\bar{\eta}\frac{L^2-D^2}{(2L^2-3D^2)S-2L},\\
T_r^L&=\frac{32}{3}\pi\bar{\eta}\frac{(L^2-D^2)D^2}{2L-D^2S},\\
T_r^D&=\frac{32}{3}\pi\bar{\eta}\frac{L^4-D^4}{(2L^2-D^2)S-2L},
\intertext{where}
S&=\frac{2}{\sqrt{L^2-D^2}}\log\left(\frac{L+\sqrt{L^2-D^2}}{D}\right),
\end{align*}
and $\bar{\eta}$ is the viscosity of the fluid.\\
To get the correct friction parameters for our model, we have to divide $T_t^L$ and $T_t^D$ by the mass $m$ and $T_r^L$ and $T_r^D$ by the moment of inertia $I_c$. Furthermore, we use the dynamic viscosity of water which is given by
$\bar{\eta}=0.001~g/cms$. Then we get
\begin{align*}
\gamma_t^L&=0.11~1/s, & \gamma_t^D&=0.12~1/s, & \gamma_r^L&=0.2~1/s, & \gamma_r^D&=1.36~1/s.
\end{align*}
Since our model has only one global friction parameter for the translation of the particles and one for the rotation of the particles, we have to do a best-fit approximation.\\
In this case, we are mostly interested in the angular distribution of the particles. Therefore we take fixed values of $\gamma$  and $A$, and fit the parameters for the rotation.\\
For the translational friction parameter $\gamma$, we take the mean between  $\gamma_t^L$ and $\gamma_t^D$ and assume no stochastic force. Hence, we set
\begin{align*}
 \gamma&=0.115~1/s, & A&=0.
\end{align*}  
For the interaction potential we choose a large strength $\epsilon_0=1000$, since the particles are solid and should not go through each other. \\
Now, the remaining parameters for the rotation are $\bar{\gamma}$ and $B$, which we want to find via fitting to the experimental data.
\subsection{Fitting Parameters}
\begin{figure}[htbp]
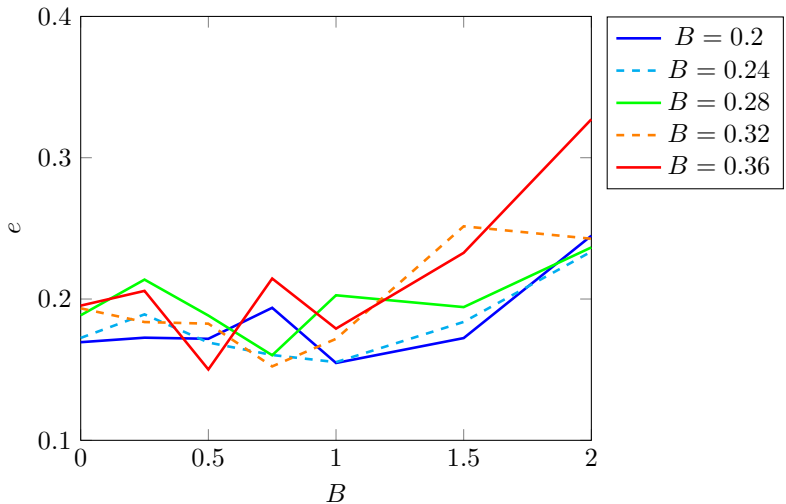

    \centering
    \resizebox{0.8\textwidth}{!}{
    \externaltikz{ExternalTikzFigure_para_fit}{
    	\relinput{Pictures/tikz/para_fit/para_fit.tex}
    }}
\caption{Relative $L^1$-error of the angular distribution for different $\bar{\gamma}$ and $B$.} 
\label{fig:para_fit}   
\end{figure}

To find the best parameter values for $\bar{\gamma}$ and $B$, we simulate the experiment for several values of $\bar{\gamma}$ and $B$ and compare the resulting angular distributions with the angular distribution of the experimental data. Therefore we choose initially $N=25$ particles equally-distributed inside $\Omega$ with equally-distributed initial angular orientation. Since in the experiment the particles are already moving, when the camera detection starts, we simulate first up to $T=5$ and use the configuration of the particles at $T=5$ as initial conditions for our simulation.\\
In the experiment the particles are detected inside of
\begin{align*}
\Omega_H=[2.5~26.5]\times[0~5]\setminus[2.5~26.5]\times[2.25~2.75].
\end{align*}
Then we derive the different angular distribution histograms with $N_H=18$ bars. Since we cannot distinguish the head and tail of a particle, we count the number of particles inside a range of 10 degrees modulus 180 degrees. This yields 18 bars, which we can compare. Let $h_i,~i=1,\ldots,N_H$ denotes the $i$-th bar value of the angular distribution histogram of the simulation for the whole domain and $h^{exp}_i,~i=1,\ldots,N_H$, the
$i$-th bar value of the angular distribution histogram of the experimental data for the whole domain, respectively. 
Then the relative error $e$ between for the histograms is given by
\begin{align*}
e=\frac{\sum_{i=1}^{18} \vert h_i-h^{exp}_i \vert}{\sum_{i=1}^{18}\vert h^{exp}_i \vert}.
\end{align*}
Now we choose $\bar{\gamma}=\{0.2,~0.24,~0.28,~0.32,~0.36\}$ and $B=\{0,~0.25,~0.5,~0.75,~1,~1.5,~2\}$
and compute the relative $L^1$-error for all different combinations. The results are shown in Figure \ref{fig:para_fit}. We observe that the best fitting angular distribution is given by the parameters $\bar{\gamma}=0.36$ and $B=0.5$. Now, we additionally compute the error of the standard deviation. With the experimental standard deviation of $\sigma^{exp}=0.997$ and the standard deviation of the simulation for the best fitting parameters $\sigma=0.987$, we get
\begin{align*}
e_{\sigma}=\frac{\vert \sigma^{exp}-\sigma \vert }{\vert \sigma^{exp}\vert}=0.01.
\end{align*}

\section{Comparison}
\label{sec:comparison}
\begin{figure}[htbp]
   \centering
   \begin{minipage}[c]{0.49\textwidth}
   \resizebox{1\textwidth}{!}{
   	\includegraphics{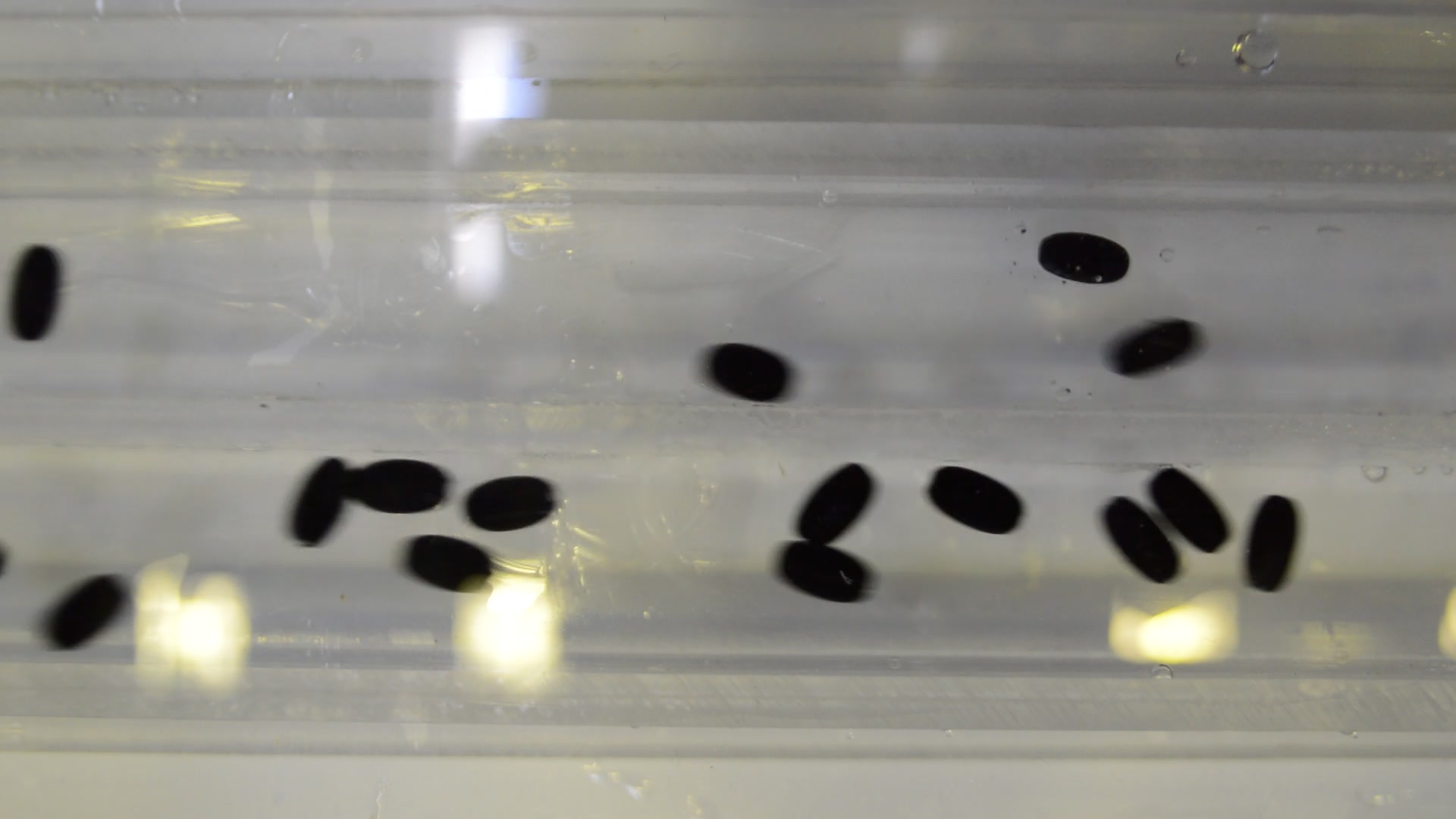}
   }
   \end{minipage}
   \begin{minipage}[c]{0.49\textwidth}
   \resizebox{1\textwidth}{!}{
      \includegraphics{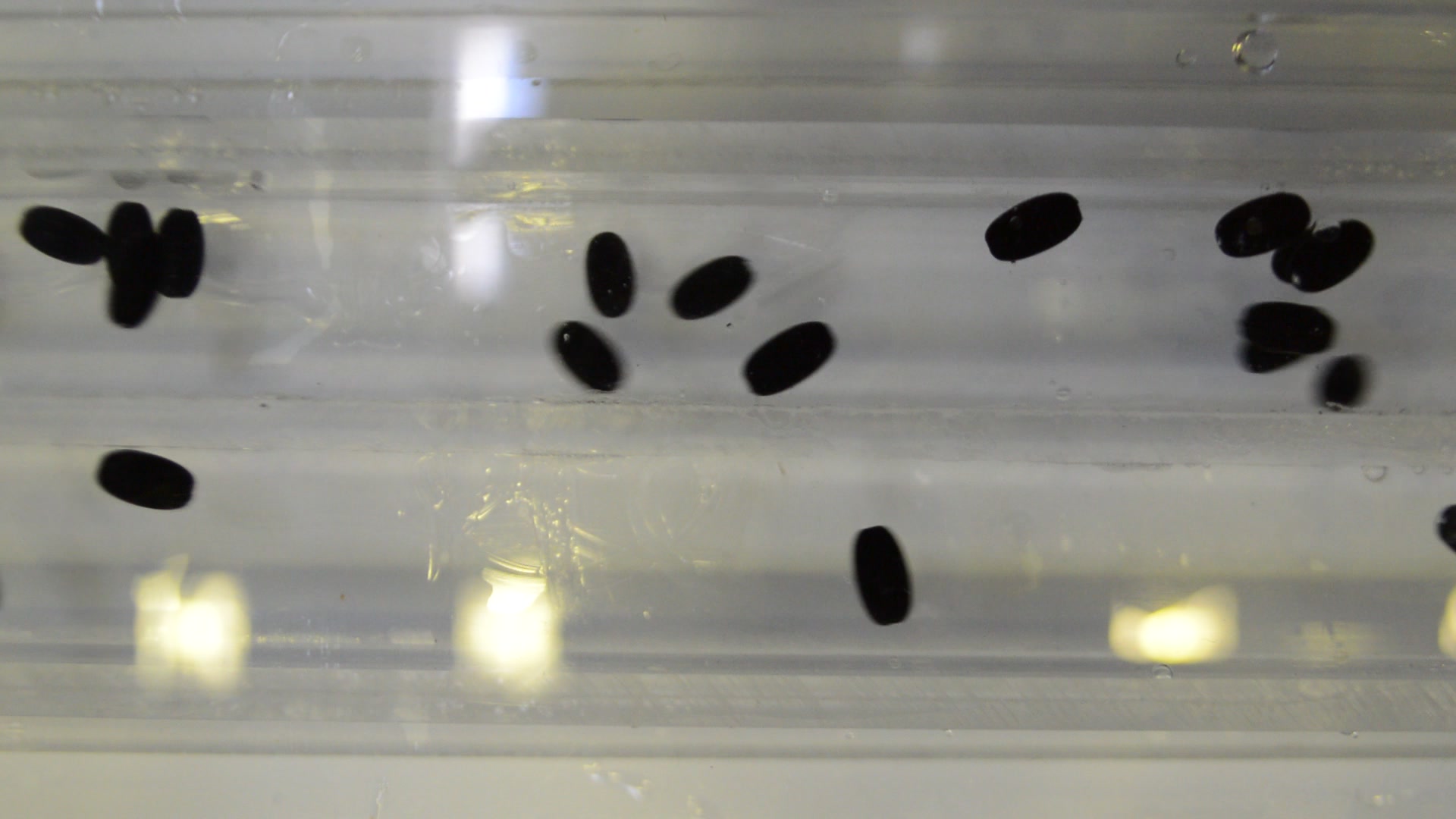}
      }
   \end{minipage}
   \begin{minipage}[c]{0.49\textwidth}
   \resizebox{1\textwidth}{!}{
   \includegraphics{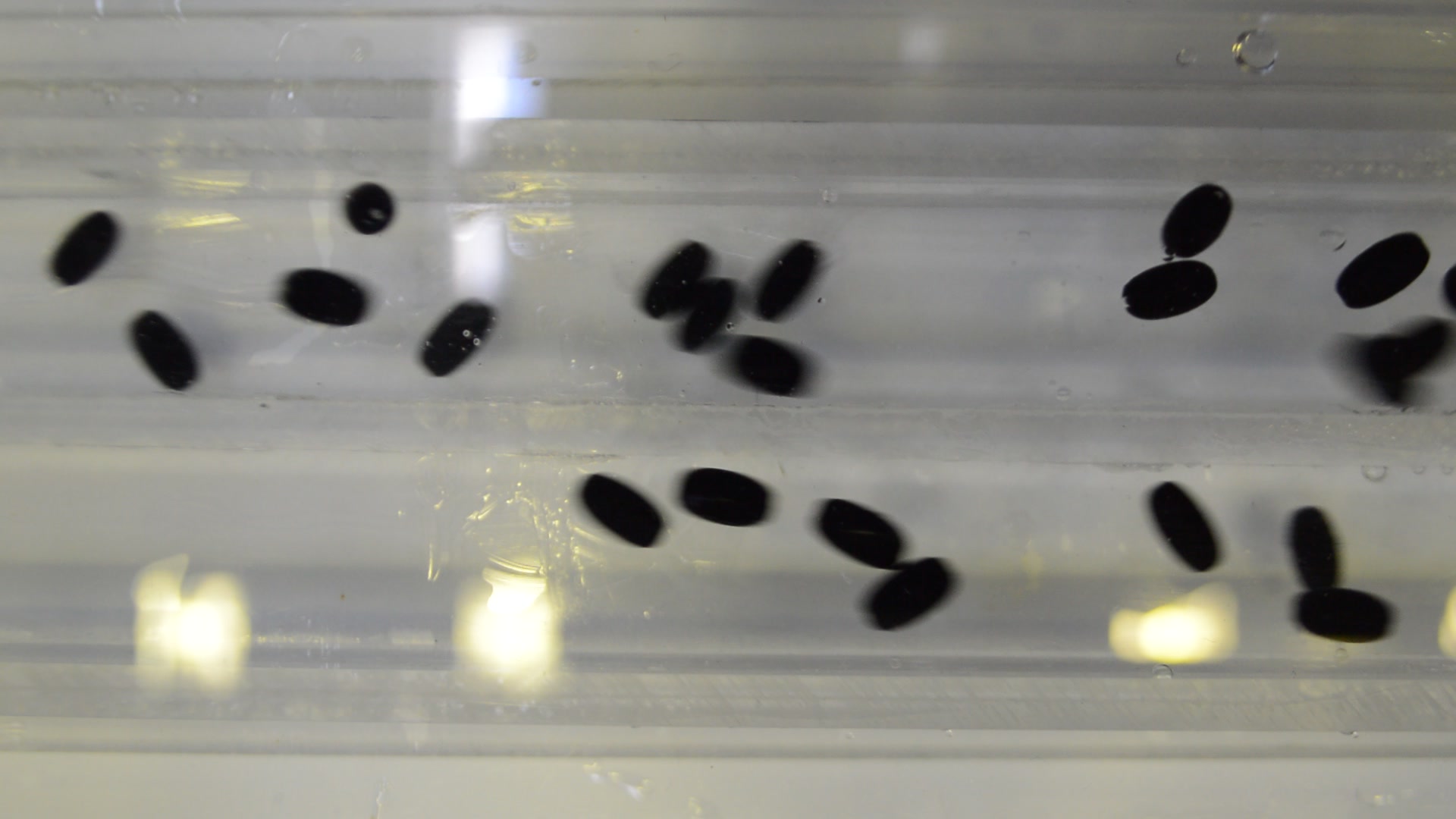}
   }
   \end{minipage}
   \begin{minipage}[c]{0.49\textwidth}
   \resizebox{1\textwidth}{!}{
      \includegraphics{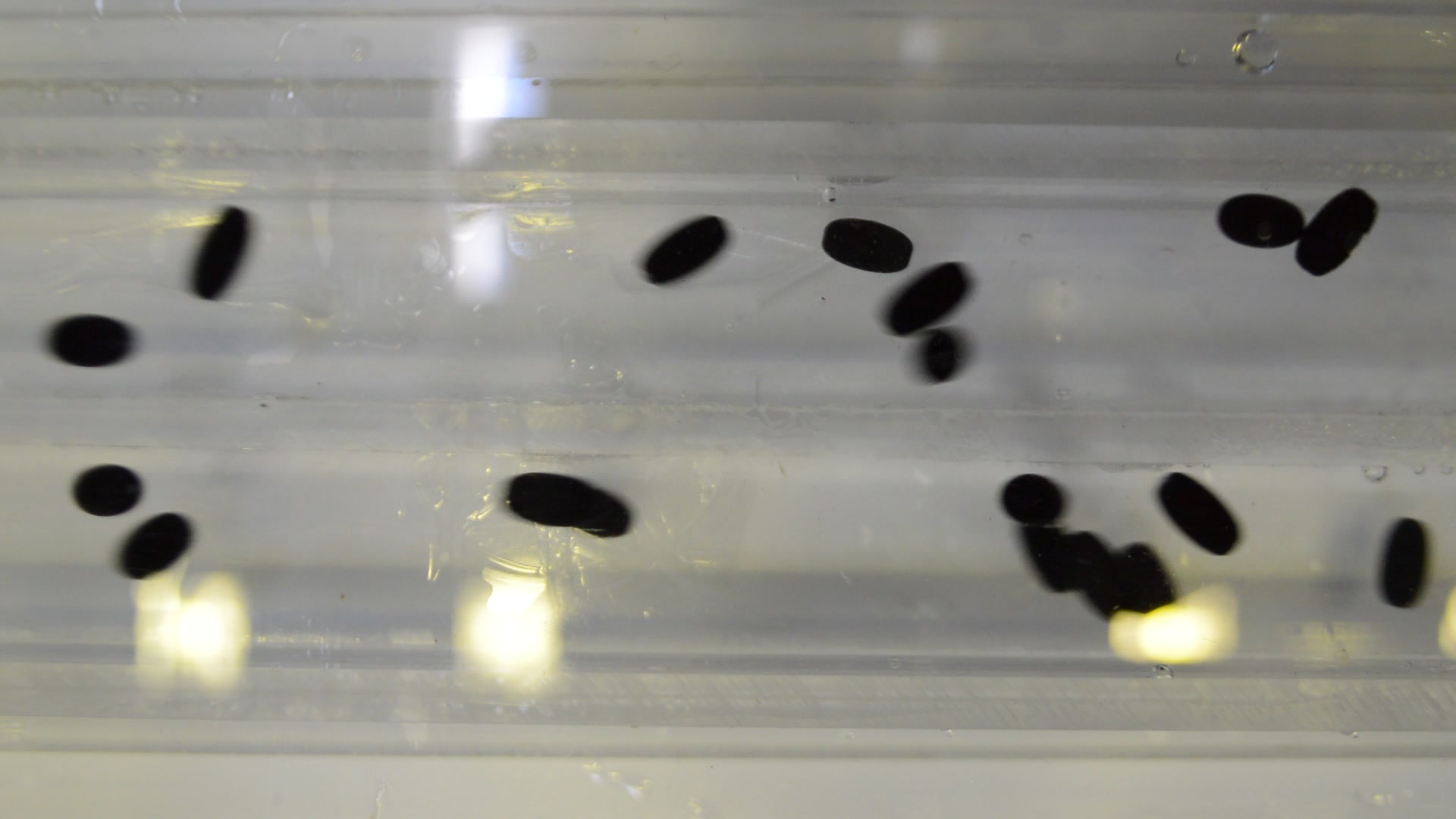}
      }
   \end{minipage}
   \caption{Pictures of the experiment with ellipsoidal particles (black particles) at times t=0, 2.5, 7.5,10.}
   \label{fig:exp_bilder}
\end{figure}

\begin{figure}[htbp]
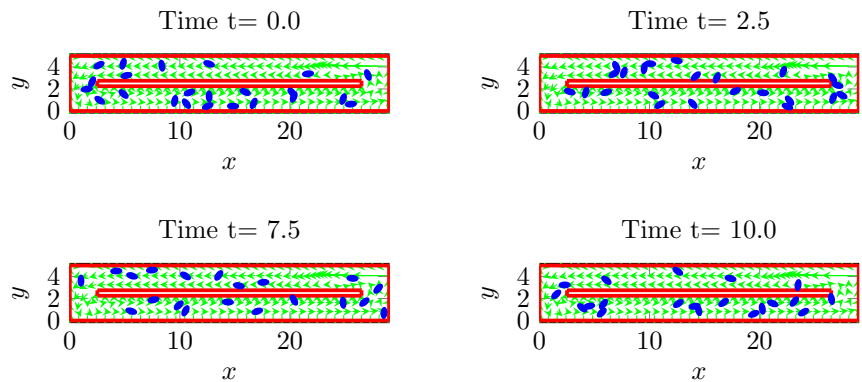

    \centering
    \externaltikz{ExternalTikzFigure_ell_exp}{
    	\relinput{Pictures/Tikz/ell_exp/ell_exp.tex}
    }
    \caption{Numerical results of \eqref{eq:micro} of the experiment for different times. The wall boundaries are indicated by the red lines. The velocity field of the surrounding fluid is given by the green arrows, and the particles are marked by blue ellipses.}
    \label{fig:exp_simulation}
 \end{figure}

\begin{figure}[htbp]
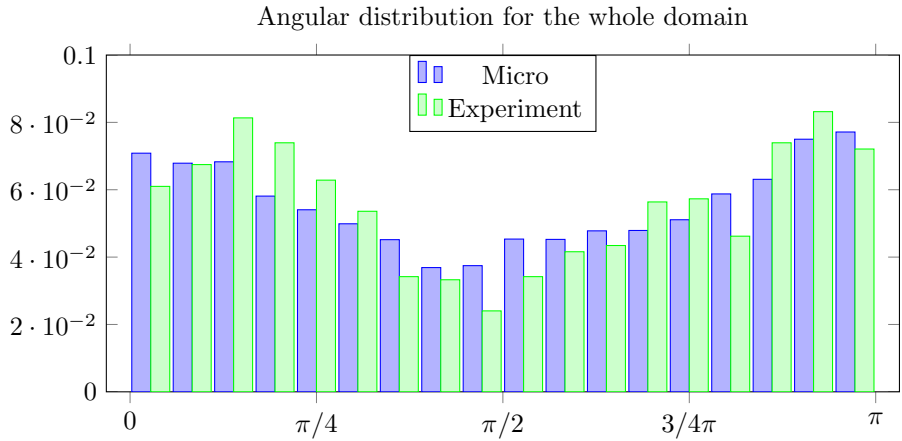

    \centering
    \externaltikz{ExternalTikzFigure_ell_exp_winkel_gesamt}{
    	\relinput{Pictures/Tikz/ell_exp_winkel_gesamt/ell_exp_winkel_gesamt.tex}
    }
    \caption{Angular distribution of the experiment for the experimental data and the numerical results of \eqref{eq:micro}.}
    \label{fig:exp_winkel}
 \end{figure}

Now we compare the experimental results with the numerical simulation. First of all, we show the position of the particles at different times of the experiment.
We observe that most of the particles orientate longitudinal to the stream of the flow, and follow the surrounding fluid (see Figure \ref{fig:exp_bilder}).
 For the numerical simulation we choose the fitting parameters $\bar{\gamma}=0.36$ and $B=0.5$. Then we also observe that the particles follow the stream of the flow and mostly orientate longitudinal to the flow (see Figure \ref{fig:exp_simulation}). Of course the particles do not have the same position as in the experiment, since their initial conditions are not exactly the same.
Now we investigate the orientation of the particles for the whole simulation. Therefore, we use the histogram results (compare Figure \ref{fig:exp_winkel}). We observe that the orientation of the particles is similar for the experiment as for the simulation. The particles have a tendency to orientate longitudinal to the stream of the flow, i.e. they orientate towards $\theta_t=0$ and $\theta_t=\pi$.
\section{Concluding Remarks}
To summarize, we can say that our ellipsoidal particle model really covers the behavior of the particles in the experimental set up. Although our model has just two space dimensions, we observe a good agreement with the experimental data, with a relative error of the standard derivation of $e_\sigma=0.01$, for the best fitting parameters. To fit also the translational parameters, more experimental data is needed, but the analysis of the parameters can be done in a similar way as the analysis presented here for the rotational parameters.\\
Furthermore, for more particles inside the channel, a macroscopic description as presented in \cite{borsche2015mean} can be investigated and also compared to experimental data, this could enable large-scale simulations with moderate computational effort.
\section*{Acknowledgment}
Funding by the Deutsche Forschungsgemeinschaft (DFG) within the RTG GrK 1932 "Stochastic Models for 
Innovations in the Engineering Sciences", project area P1, is gratefully acknowledged.

\bibliographystyle{elsarticle-num}  
\bibliography{literatur}

\end{document}